\documentclass[pra,twocolumn,aps,showpacs,preprintnumbers,amsmath,amssymb,superscriptaddress]{revtex4-1}

\usepackage{graphicx}
\usepackage{dcolumn}
\usepackage{bm}
\usepackage{amsmath}
\usepackage{cancel}

\begin{document}


\title{Generalized Grover's algorithm for multiple phase inversion states}

\author{Tim Byrnes}
\affiliation{State Key Laboratory of Precision Spectroscopy, School of Physical and Material Sciences,
East China Normal University, Shanghai 200062, China}
\affiliation{New York University Shanghai, 1555 Century Ave, Pudong, Shanghai 200122, China}
\affiliation{NYU-ECNU Institute of Physics at NYU Shanghai, 3663 Zhongshan Road North, Shanghai 200062, China}
\affiliation{National Institute of Informatics, 2-1-2 Hitotsubashi, Chiyoda-ku, Tokyo 101-8430, Japan}
\affiliation{Department of Physics, New York University, New York, NY 10003, USA}

\author{Gary Forster}
\affiliation{Department of Physics, University of Bath, Bath BA2 7AY, UK}
\affiliation{National Institute of Informatics, 2-1-2 Hitotsubashi, Chiyoda-ku, Tokyo 101-8430, Japan}

\author{Louis Tessler}
\affiliation{New York University Shanghai, 1555 Century Ave, Pudong, Shanghai 200122, China}
\affiliation{CEMS, RIKEN, Wako-shi, Saitama 351-0198, Japan}

\date{\today}

\begin{abstract}
Grover's algorithm is a quantum search algorithm that proceeds by repeated applications of the Grover operator and the Oracle until the state evolves to one of the target states.  In the standard version of the algorithm, the Grover operator inverts the sign on only one state.  Here we provide an exact solution to the problem of performing Grover's search where the Grover operator inverts the sign on $ N $ states.  We show the 
underlying structure in terms of the eigenspectrum of the generalized Hamiltonian, and derive an appropriate initial state to perform the Grover evolution.  This allows us to use the quantum phase estimation algorithm to solve the search problem in this generalized case, completely bypassing the Grover algorithm altogether.  We obtain a time complexity of this case of $ \sqrt{D/M^\alpha} $ where $ D $ is the search space dimension, $ M $ is the number of target states, and $ \alpha \approx 1 $, which is close to the optimal scaling. 
\end{abstract}

\maketitle

Grover's algorithm \cite{grover1996} is one of the central algorithms in the field of quantum computing that shows a speedup in comparison to classical computing. For an unsorted search space with $ D $ elements, classical algorithms take $ \propto D $ steps to find a solution, in comparison to Grover's algorithm taking $ \propto \sqrt{D} $ steps.  While the speedup is only quadratic in comparison to other quantum algorithms such as Shor's algorithm with an exponential speedup, it is of fundamental interest as it 
can be applied to very wide variety of problems. 
Many variants and applications of Grover's algorithm have been investigated in the past.  The concept of searching can be generalized to abstract solution spaces rather than literal databases, making it applicable in principle to any NP problem \cite{Bennett1997,Furer2008}. Furthermore Grover search finds many uses as a primitive in diverse applications such as cryptography \cite{Hsu03,Hao2010}, 
matrix and graph problems \cite{Magniez07,5954250}, quantum control tasks \cite{Chen05}, optimization \cite{Durr96aquantum, Grover1997}, element distinctness \cite{2005quant.ph.4012A}, collision problems \cite{Brassard1997}, and quantum machine learning \cite{Aimeur2013}.

\begin{figure}
\includegraphics[width=\columnwidth]{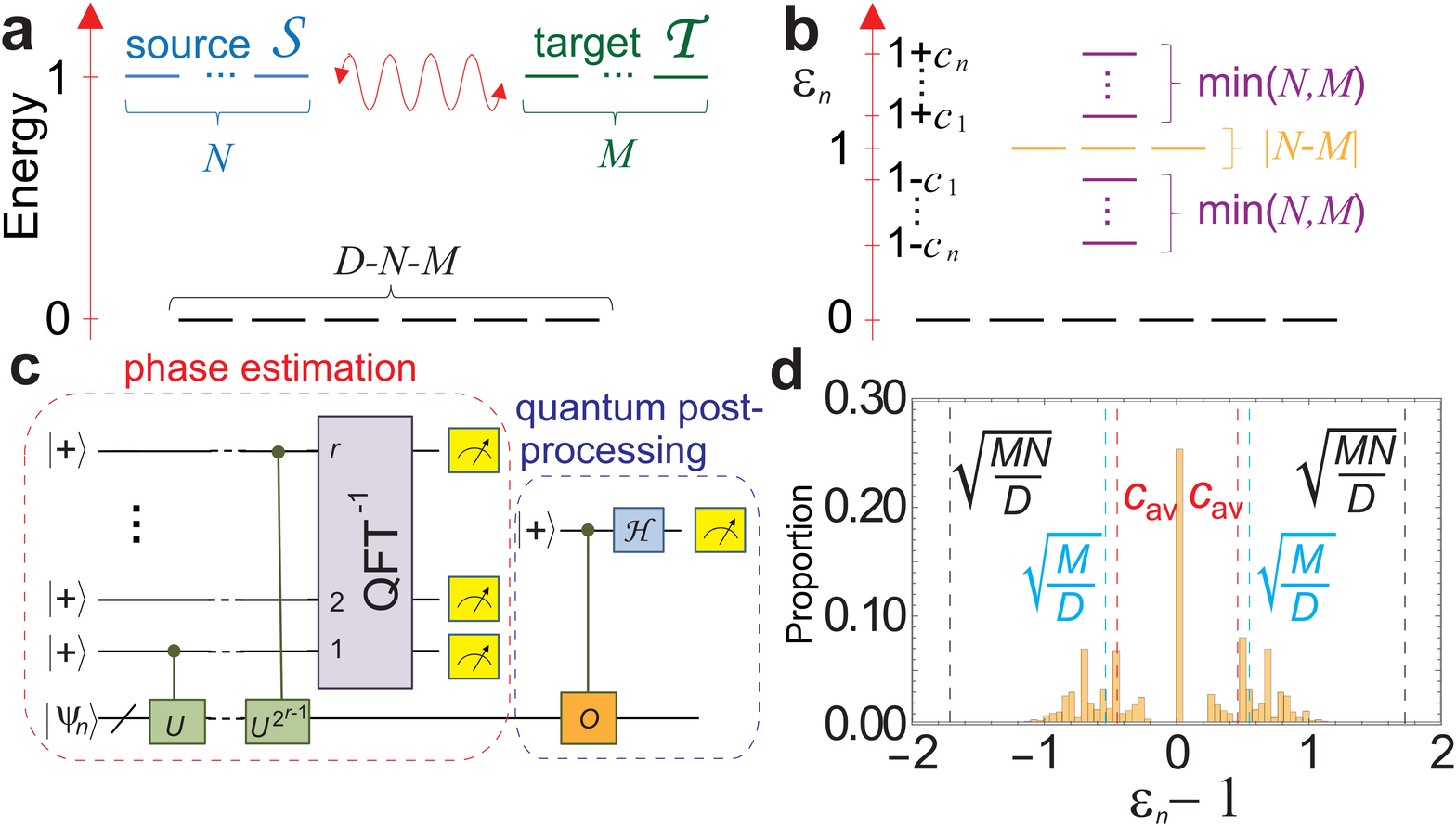}
\caption{(a) Interpretation of the generalized Grover evolution as Rabi oscillations between source $ \cal S $ and target $ \cal T $ subspaces.  (b) Energy spectrum of the Grover Hamiltonian after diagonalization.  States in the source and target sector appear in pairs with energy $ \epsilon^\pm_n = 1 \pm |c_n| $.  Unpaired states in $ {\cal S } $ and $ \cal T $ have an energy of 1, and all remaining states have energy 0. (c) Quantum circuit which produces a state in the target sector $ \cal T $ for the generalized Grover algorithm.  Here $ U = e^{-iH} $, where $ H $ is (\ref{firstham}), $ O $ is the Oracle, and $ \cal H $ is a Hadamard gate, and $ \text{QFT}^{-1} $ is an inverse quantum Fourier transform. (d) Distribution of eigenvalues of the Grover Hamiltonian (\ref{firstham}) for initial states of the form  $ | \psi_n \rangle = {\cal H} | n \rangle $ and $ N = M = 10 $ and $ D =2^5 $. The average value of $ |c_n| $ over all choices of initial state  $ c_{\text{av}} $ is compared to the standard Grover scaling of $ \sqrt{M/D} $ and the upper bound $ \sqrt{MN/D} $.  }
\label{fig1}
\end{figure}

The standard version of Grover's algorithm proceeds by first preparing the register in a equal superposition of all states $ | + \rangle = \frac{1}{\sqrt{D}} \sum_{n=0}^{D-1} | n \rangle $. One then repetitively applies the Oracle operator $ O = I - 2 \sum_{n \in {\cal T}} | n \rangle \langle n | $
where $ {\cal T} $ is the set of target (i.e. solution) states, and the Grover operator $ G_0  = I - 2 | 0 \rangle \langle 0 |  $, interspersed with Hadamard operations. The Hadamard operations can be combined with the $ G_0 $ by defining $ G = I -  2 | + \rangle \langle + | $
such that for $ \frac{\pi}{4} \sqrt{\frac{D}{M}}$ applications of $ G O $ gives with high probability a target state \cite{nielsen00}. There is an obvious asymmetry between the operators $ G $ and $ O $, as the Oracle inverts the phase of multiple target states, while the Grover operator only inverts the sign of one state.  The generalization where both $ G $ and $ O $ inverts the phase on multiple states was previously studied by Sadhukhan and Tulsi \cite{sadhukhan12}.  In their work an analytic solution was found for $ N =2 $ and $ M =2 $, where $ N $ is the number of states that the Grover operator inverts and $ M $ is the number of target states.  However, for larger $ N, M $ only numerical solutions could be obtained. Another generalization was performed by Kato \cite{kato05} where the Grover operator was modified to one with a Hamiltonian only including single qubit operators.  This corresponds to a different situation where a more general phase (not just $ \pm 1 $) are put on a spectrum of states by the Grover operator.  The algorithm works in an asymptotic sense where the number of qubits is large.  Other generalizations of Grover's algorithm such as for continuous evolution \cite{PhysRevA.57.2403}, zero failure rate \cite{PhysRevA.64.022307}, arbitrary initial amplitude distribution \cite{Biron:1998ic}, and fixed-point search \cite{grover2005fixed,yoder2014fixed} have been investigated.  To our knowledge, a general solution to the case of solving the Grover problem for arbitrary $ N, M $ is not currently available.

The problem of generalizing to any $ N, M $ is of interest in situations where no simple physical implementation is available to perform  $ G $ simply. For example, in continuous variable formulations of quantum computing \cite{braunstein05,byrnes2014}, it may be impractical or undesirable to only put the phase on a single quantum space in an infinite Hilbert space \cite{pati00,byrnes2012,PhysRevA.66.042310}. The Grover operator in this case would correspond to inverting the phase on an infinitely squeezed momentum state, which may be difficult to achieve in practice and also has a vanishing overlap with solution states encoded in position eigenstates. As we describe in this paper, the case with arbitrary $ N, M $ gives a more general formulation of the problem, as a population transfer between two subspaces of a larger Hilbert space.  It can also lead to a reduction of resources by a simpler implementation of the Grover operator. To perform a phase flip on a single state requires a multi-qubit controlled-$Z$ gate which is decomposable to elementary gates that grow as the square of the number of qubits \cite{nielsen00}.  We show also that it is possible to apply the quantum phase estimation algorithm in order to perform the Grover search, and bypass Grover's algorithm altogether.  This suggests interesting implications for the classifications of quantum algorithms, in view of the fact that amplitude amplification and phase estimation are usually considered to be distinct roots of the dependency tree for quantum algorithms \cite{nielsen00}.  We also note that our framework allows for the opportunity to apply our scheme as a subroutine in other quantum algorithms that use related methods \cite{terhal98,poulin09,farhi2014quantum}.

We show our generalization first for the continuous time version of the Grover algorithm, where a single Hamiltonian evolves the state from the initial state to the target states \cite{nielsen00,PhysRevA.57.2403} (see Supplementary Information).   The generalized Grover Hamiltonian reads
\begin{align}
H & = P_{\cal S} + P_{\cal T} 
\label{firstham}
\end{align}
where $ P_{\cal S} \equiv \sum_{n \in {\cal S}} | \psi_n \rangle \langle \psi_n |  $ and $ P_{\cal T} \equiv \sum_{n \in {\cal T}} | n \rangle \langle n | $ are projection operators for the space of states as defined by the source $ {\cal S} $ and target $ {\cal T} $ respectively.  Here, the parameters $ N = | {\cal S} | $ and $ M = | {\cal T} | $ correspond to the rank of the projectors $ P_{\cal S} $ and $ P_{\cal T}  $ respectively. We have also made the generalization that the states in the target and source states are of arbitrary form, except for orthogonality $ \langle \psi_n| \psi_{n'} \rangle  = \delta_{nn'} $ and $ \langle n | n' \rangle  = \delta_{n n'} $.  We assume that the source states are not orthogonal to the target space $ \langle \psi_n | P_{\cal T}  | \psi_n \rangle > 0 $ and the rank of 
$ H $ is $ N+M $ such that the source and target subspaces do not contain each other $ P_{\cal S} P_{\cal T} \ne P_{\cal S}, P_{\cal T} $.

There is an intuitive way to understand the Hamiltonian formulation of Grover's algorithm as Rabi oscillations between the source and target subspaces. Viewing (\ref{firstham}) in energy space, the effect of the Grover Hamiltonian is to specify particular states (those in $ \cal S $ and $ \cal T $) in the Hilbert space to have an energy of 1, which implicitly sets all the remaining states to have an energy 0  (Fig. \ref{fig1}(a)).   Since the states in $ \cal S $ and $ \cal T $ are not mutually orthogonal, there is a transition matrix element between them equal to the overlap between the states (see Supplementary Information).  The time complexity in this formulation originates from the need to evolve the Hamiltonian from the initial to final state, which is the time required for half a Rabi oscillation. For $ N =1 $ the overlap between $ |+ \rangle $ and the superposition state over all $ \cal T $ is $ \sqrt{M/D} $.  The time for the Rabi oscillation is then proportional to inverse of this (working in units $ \hbar = 1 $), giving a scaling $ \propto  \sqrt{D/M} $.

\begin{figure}[t]
\includegraphics[width=\columnwidth]{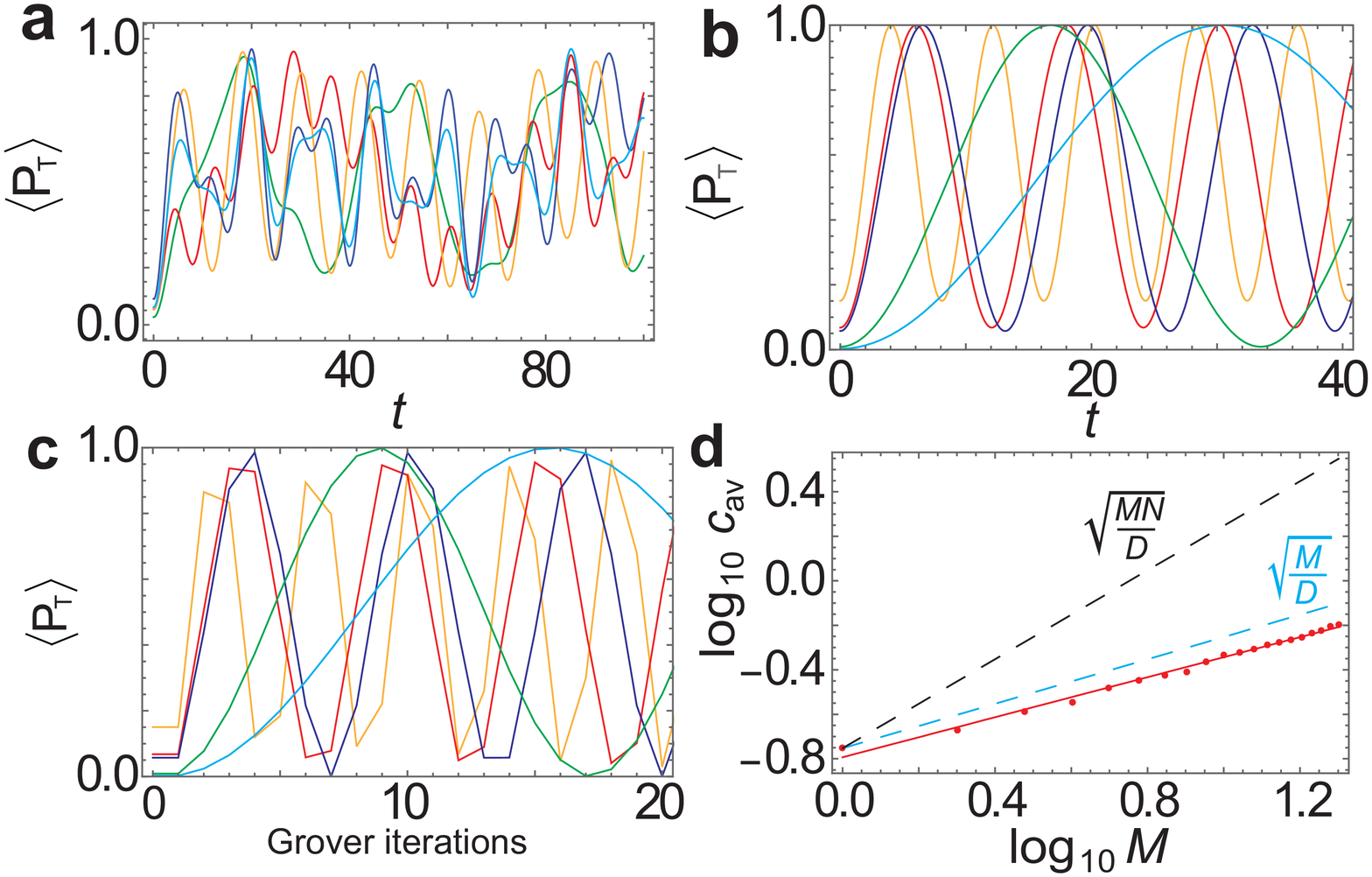}
\caption{(a) Time evolution with the generalized Grover Hamiltonian with various initial states chosen as $ | \psi_n \rangle $.  (b) Time evolution choosing various initial state $ | \Psi_n(t=0) \rangle $.  (c) Evolving the various $ | \Psi_n(t=0) \rangle $ using a gate based Grover iteration.  One Grover iteration corresponds to  the combined application of  $ G =  e^{-i \pi P_{\cal S}} = 1 - 2 P_{\cal S}$ and $ O = e^{-i \pi P_{\cal T}} = 1- 2 P_{\cal T} $. For (a)(b)(c) $ D = 100 $, $ N = 5 $ source states and $ M = 5 $ target states. The source states  $ | \psi_n \rangle $ are taken to be orthonormal random vectors. (d) Scaling of the average energy separation $ c_{\text{av}} $ for $ M = N $, $ D = 2^5 $, and averaged over random choices of
$ | \psi_n \rangle = {\cal H} | n \rangle $ (points).  Scaling of the maximum $ |c_n |$ ($\propto \sqrt{MN/D} $) and standard Grover result ($\propto \sqrt{M/D} $) are shown for comparison (dashed lines). A straight line fit of the points gives a slope of $ \alpha/2 \approx 0.45 $.  }
\label{fig2}
\end{figure}

If $ \cal S $ contains more than one state $ N > 1 $, simply preparing 
the state in one of the source states $ | \psi_n \rangle $ does not produce clean oscillations. In Fig. \ref{fig2}(a) an example of this is shown, where initial states are chosen to be the same as the source states.  For any case with $ N > 1 $ the time evolution fails to give predictable oscillations. Furthermore, the probability of reaching the target sector tends to diminish with $ N $.  Without clean oscillations the algorithm is difficult to handle as it is hard to predict what time to evolve the Grover Hamiltonian, and the success probability is also reduced.  This can however be remedied by choosing a suitable initial state as we show below.  

The Hamiltonian (\ref{firstham}) has special properties which can be exploited for the case $ N,M> 1 $. Split the Hamiltonian into two subspaces, defined by states spanned by the states in $ \cal T $ (dimension $M \times M $) and all the remaining states $ \cancel{\cal T} $ (dimension $ D-M \times D-M$).  Defining $  P_{\cancel{\cal T}} \equiv 1 -  P_{\cal T} $, the Hamiltonian can then be written
\begin{align}
H = (P_{ \cancel{\cal T}} + P_{\cal T}) H (P_{ \cancel{\cal T}} + P_{\cal T}) 
= \left(
\begin{array}{cc}
A & B \\
B^\dagger & C 
\end{array}
\right)
\end{align}
where the submatrices are defined as $ A \equiv P_{ \cancel{\cal T}} P_{\cal S}  P_{ \cancel{\cal T}} , B \equiv P_{ \cancel{\cal T}} P_{\cal S}  P_{ \cal T} , C \equiv   P_{ \cal T}  P_{\cal S} P_{ \cal T} + P_{ \cal T} $.  Here, $ A $ and $ C $ are Hermitian.  Due to the special form of the submatrices above, we now show that diagonalizing $ A $ and $ C $ simultaneously diagonalizes $ B $.  To see this, we may use the standard properties of the projection operators to show
\begin{align}
B B^\dagger & = A - A^2 
\label{matrixrelation1} \\
B^\dagger B & = -C^2 + 3 C -2 P_{\cal T} .
\label{matrixrelation}
\end{align}
It thus follows that  $ [ B B^\dagger , A ] = [ B^\dagger B , C ]  = 0  $
so that $ B B^\dagger $ and $ A $ share the same eigenvectors, and similarly for $ B^\dagger  B $ and $ C $. The matrices can be written $ A = U_{\cancel{\cal T}} \Lambda_A U_{ \cancel{\cal T}}^\dagger$, $ B = U_{ \cancel{\cal T}} \Lambda_B U_{ \cal T}^\dagger $, and $ C = U_{\cal T} \Lambda_C U_{\cal T}^\dagger $, in terms of their diagonalized matrices $ \Lambda $ and $ U_{ {\cal T} } $, $ U_{ \cancel{\cal T}} $  are unitary rotations in the spaces $ { \cal T} $, $ \cancel{\cal T} $ respectively.  
Eq. (\ref{matrixrelation1}) and (\ref{matrixrelation}) allows us to deduce the relationship between the eigenvalues of the matrices.  Let us write the eigenvalues of the matrix $ C $ as
\begin{align}
(\Lambda_C)_{nn'} = (1 + |c_n|^2) \delta_{nn'} .
\label{deigenvalues}
\end{align}
where we used the fact that $ P_{ \cal T}  P_{\cal S} P_{ \cal T} $ is positive definite to write its eigenvalue is $ |c_n|^2 $, and $ 1 \le n \le M $ here as $ C $ is of rank $ M $. 
Substituting this into (\ref{matrixrelation}) we may deduce that the eigenvalues of $ B $ are
\begin{align}
(\Lambda_B)_{nn'} =  \delta_{nn} c_n \sqrt{ 1- |c_n|^2}
\end{align}
This may be in turn be used in the quadratic equation (\ref{matrixrelation1}) to deduce that the eigenvalues of $ A $ are of two types: $ (\Lambda_A)_{nn} = 1 - |c_n|^2 ,  |c_n|^2 $.  We also require consistency with the property of the Hamiltonian $ \text{Tr} ( H) = N+M $, which should be invariant under unitary transformations. The eigenvalue type $  1 - |c_n|^2 $ combined with (\ref{deigenvalues}) ensures this consistency.  The remaining eigenvalues are of the second type with $ |c_n|^2 =0 $, so that
\begin{align}
(\Lambda_A)_{nn'} = \left\{
\begin{array}{cc}
(1 - |c_n|^2 ) \delta_{nn'} & 1 \le n \le N  \\
0 & \text{otherwise}
\end{array}
\right. .
\label{aeigenvalues}
\end{align}
In order that (\ref{deigenvalues}) and (\ref{aeigenvalues}) give $ \text{Tr} ( H) = N+M $, there can be then at most $ \min(N,M) $ of the $ |c_n|^2 $ to be nonzero.  

With the rotation of only $ U_{ \cal T} $ and $ U_{ \cancel{\cal T}} $, the  Hamiltonian may therefore be put in $ 2 \times 2 $ block diagonal form
\begin{align}
& H  = \sum_{n=1}^{\min (N,M)} \Big[  (1 - |c_n|^2) | \epsilon_n^{  \cancel{\cal T} } \rangle \langle \epsilon_n^{  \cancel{\cal T} } | 
 + c_n \sqrt{ 1- |c_n|^2}  | \epsilon_n^{  \cancel{\cal T} } \rangle \langle \epsilon_n^{  \cal T } | \nonumber \\
& + c_n^* \sqrt{ 1- |c_n|^2}  | \epsilon_n^{  \cal T } \rangle \langle \epsilon_n^{  \cancel{\cal T} }  | 
+  (1 + |c_n|^2) | \epsilon_n^{  \cal T } \rangle \langle \epsilon_n^{  \cal T}  |  \Big] \nonumber \\
& + \sum_{n=\min (N,M)+1}^{\max (N,M)} \Big[  \theta_{N-M} | \epsilon_n^{  \cancel{\cal T} } \rangle \langle \epsilon_n^{  \cancel{\cal T} } | + \theta_{M-N} | \epsilon_n^{  \cal T } \rangle \langle \epsilon_n^{  \cal T}  | \Big] ,\label{hamblock} 
\end{align}
where $ | \epsilon_n^{  \cancel{\cal T} } \rangle, | \epsilon_n^{ \cal T} \rangle  $ are the eigenvectors for the $ A $ and $ C $ matrices respectively, and $ \theta_m = 1 $ for $ m > 0 $ and zero otherwise. We emphasize that the fact that $ B $ diagonalizes here is nontrivial, without which we would not have the simple $ 2 \times 2 $ block diagonal structure. 
\begin{align}
| \epsilon_n^{\pm} \rangle = \sqrt{\frac{1 \mp |c_n|}{2}}  | \epsilon_n^{  \cancel{\cal T} } \rangle \pm \sqrt{\frac{1 \pm |c_n|}{2}}     | \epsilon_n^{  \cal T } \rangle
\label{eigenvectors}
\end{align}
for $ 1 \le n \le \min (N,M) $ with eigenvalues
\begin{align}
\epsilon_n^{\pm} = 1 \pm |c_n| .
\label{hameigen}
\end{align}
The remaining $ N+M-2 \min (N,M) = |N-M| $ eigenvalues all are 1, which corresponds to having $ c_n = 0 $.   We thus obtain a diagonalized energy spectrum of the form shown in Fig. \ref{fig1}(b), where the nontrivial eigenvalues are arranged in pairs centered around an energy 1, and the remaining at exactly 1.

%

For the purposes of solving the search problem,
\begin{align}
| \epsilon_n^{  \cal T} \rangle = \sqrt{\frac{1+ |c_n|}{2}} | \epsilon_n^{+} \rangle -  \sqrt{\frac{1- |c_n|}{2}}  | \epsilon_n^{-} \rangle  
\label{finalstate}
\end{align}
is precisely the desired vector since it is by definition a state which is completely in the target space. This can be achieved by preparing 
\begin{align}
|\Psi_n (t=0) \rangle =  \sqrt{\frac{1+ |c_n|}{2}} | \epsilon_n^{+} \rangle +  \sqrt{\frac{1- |c_n|}{2}}  | \epsilon_n^{-} \rangle 
\label{initialstate}
\end{align}
and time-evolving this state under the Grover Hamiltonian until a relative minus sign is picked between the two terms.  This occurs at a $ t = \pi/2|c_n| $ as the state $ | \epsilon_n^{\pm} \rangle $ has a time evolving phase of $ e^{-i(1 \pm |c_n|)t } $ according to (\ref{hameigen}). We numerically confirm that perfect Grover oscillations are achieved if the state (\ref{initialstate}) is prepared for any $ N, M $ and evolved under the Grover Hamiltonian.    In Fig. \ref{fig2}(b) we see that the oscillations take a perfect sinusodial form, with the probability of reaching the target subspace reaching 1 at times $ t = \pi/2|c_n| $.  Although derived for the Hamiltonian formulation of Grover's algorithm, the initial state (\ref{initialstate}) also works for the gate based version of Grover's algorithm, where the signs are inverted on the source and target states in sequence.  Fig.  \ref{fig2}(c) shows the evolution under such Grover iterations for the same choice of random source states.  The evolution shows a similarity to Fig.  \ref{fig2}(b) which is as expected in the view that the gate version of Grover's algorithm is a Trotter expansion of the Grover Hamiltonian \cite{nielsen00}.  Some of the faster oscillations do not reach a probability 1 due to the relatively small Hilbert space of states that are used in the simulation, where it is easy to overshoot the maximum in a discrete evolution.

For the standard Grover case ($ N = 1 $), the initial state (\ref{initialstate}) takes a convenient form $ |\Psi(t=0) \rangle = | \psi_{n=1} \rangle $ independent of the target states $ \cal T $.  Unfortunately, for the $ N > 1 $ case there is no unique initial state that can be prepared that is independent of the target states. This is a serious issue, as it suggests that one requires knowledge of the matrices $ A $ and $ C $, which in turn requires knowledge of the target states in advance, defeating the purpose of the algorithm.  We however introduce an alternative procedure which is based on the phase estimation algorithm, which overcomes this problem \cite{Brassard98,mosca2001counting}. 

Instead of time-evolving the Grover Hamiltonian, we directly prepare the desired state (\ref{finalstate}) using a quantum circuit as shown in Fig. \ref{fig1}(c) (see Supplementary Information).  The algorithm involves two steps.  In the first step, phase estimation is used to obtain an eigenstate $ |\epsilon^\pm_n \rangle $  of the  Grover Hamiltonian. This can be prepared with high probability by putting any one of the source states $ | \psi_n \rangle $ as the input of the phase estimation and measuring the register.  The source states $ | \psi_n \rangle $ can be represented with high fidelity in terms of $ |\epsilon^\pm_n \rangle $, since these fully span the space $ \cal S $ as long as $ M \ge N $. Working with  $ M \ge N $ avoids the presence of the $ | \epsilon_n^{  \cancel{\cal T} } \rangle $ eigenstates in (\ref{hamblock}) which reduce the success probability, we henceforth assume this condition. Using the eigenstates $ |\epsilon^\pm_n \rangle $ as an input to the ``quantum post-processing'' (QPP) part of the circuit, which gives an output before measurement $ \sqrt{1 \mp |c_n|} | 0 \rangle | \epsilon_n^{  \cancel{\cal T} } \rangle \pm \sqrt{1 \pm |c_n|} | 1 \rangle | \epsilon_n^{  {\cal T} } \rangle $.
On measurement of the ancilla qubit, a state in the target subspace is obtained by postselecting the outcome $ |1 \rangle $.  This occurs with probability close to $1/2 $, because for a small overlap of the source and target spaces $ |c_n| \ll 1 $ \footnote{ Alternatively, a simpler procedure is to measure the output of the phase estimation algorithm (i.e. the state \unexpanded{$ |\epsilon^\pm_n \rangle $}) in the \unexpanded{$ | n \rangle $} basis and then checking with the Oracle whether the state is in the solution space.  In this case the quantum post-processing part of Fig. \ref{fig1}(c) is not required, although the ability to measure in the \unexpanded{$ | n \rangle $} basis is required.  Both versions have an approximate success probability of 1/2.}.

What is the time complexity for this phase estimation version of Grover's algorithm? The QPP only adds an constant overhead to the algorithm, hence this is negligible.  The execution time of phase estimation entirely depends upon the desired precision of the eigenvalue readout.  To perform the phase estimation, controlled-$U$ gates to the power of $ 2^{k} $ are required, where $ 0 \le k \le r-1 $, $ r $ is the number of register qubits in the phase estimation circuit, and $ U = e^{iH} $.  As there is no simplified way in general of performing the powers of $ U $, this part must be evolved directly by evolving the Grover Hamiltonian to times $ 2^{k} $. The total time of the search algorithm using the phase estimation is dominated by the number of controlled-$U$ gates, which is $ \approx \sum_{k=0}^{r-1} 2^k \approx 2^r  $.  The $ r $ required sets the energy resolution $ \delta E $ of the phase estimation readout.  
The number of register qubits required for a given energy resolution can be related according to $ \delta E $, is $ r = -\log_2 \delta E + \log_2 (2 + \frac{1}{2(1-p)} ) $ \cite{nielsen00},  where probability $ p $ of the phase estimation succeeding to classify a given state into the energy resolution. In our case, the required energy resolution is set by the energy difference between the $ |\epsilon^+_n \rangle $ and $ |\epsilon^-_n \rangle $, which is $ \epsilon^+_n - \epsilon^-_n = 2 |c_n| $. Since there are $ N $ pairs of eigenstates  $ |\epsilon^\pm_n  $, we can estimate the required energy solution as $ \delta E \le 2 c_{\text{av}} $, where the average is $ c_{\text{av}} = \sum_n |c_n| / N $.  Taking into account of the 1/2 success probability of the quantum post-processing, we finally arrive at a time scaling of the algorithm 
\begin{align}
T \approx \frac{2 + \frac{1}{2(1-p)} }{c_{\text{av}}}  .
\label{timescaling}
\end{align}

The time scaling of the algorithm depends upon the energy spectrum, which in turn depends on particular choice of states $ | \psi_n \rangle $.  For infinitesimal overlap of the source and target, the $ |c_n| $ are also infinitesimal and the time diverges.  More typically, one would choose source states that are a superposition of all states.  As an example, let us examine the case where the source states are $ | \psi_n \rangle = {\cal H} | n \rangle $ for $ n \in {\cal S} $, where $ {\cal H} $ is the Hadamard operation producing an equal amplitude superposition of all states.  The scaling of the energies can be shown to be exactly $ | c_n | \propto \frac{1}{\sqrt{D}} $, and bounded by  $ | c_n | \le \sqrt{MN/D} $.  Figure \ref{fig1}(d) shows a plot of the typical distribution of the eigenvalues $ \epsilon_n - 1 = \pm |c_n| $ of the Grover Hamiltonian. We see that the eigenvalues are bounded by the relation $ | c_n | \le \sqrt{MN/D} $ as expected, but most are distributed in a range that is much less than this.  The average $ c_{\text{av}} $ is very close to the standard Grover scaling of  $ \sqrt{M/D} $. To obtain the scaling of $ c_{\text{av}} $ with respect to $ M $, we numerically average over random choices of $ | \psi_n \rangle $, for $ N =  M$ and fixed $ D $.  We see that the scaling shows a similar exponent to the standard Grover case.  Putting this into (\ref{timescaling}) we obtain a time resource estimate for the $ N = M $ Grover's algorithm with Hadamard source states as
\begin{align}
T \propto \sqrt{\frac{D}{M^{\alpha}}}
\label{finaltimescaling}
\end{align}
where the  $ \propto \sqrt{D} $ is exact and we estimate $ \alpha \approx 0.9 $.  This is consistent with the bounds derived in Refs. \cite{boyer98,zalka99}.  Thus while it is possible for some eigenvalues $ | c_n | $ to exceed the bound, on average it is consistent with the optimal scaling of $c_{\text{av}} \propto \sqrt{M/D} $.    

In summary, we have generalized Grover's algorithm to the case where a sign inversion is performed by the Grover operator for $ N $ states and the Oracle for $ M $ states.   We find that provided the state is initialized in a suitable state (\ref{initialstate}), the time evolution of the Grover Hamiltonian induces oscillations between the source and the target sector in the same way as the standard Grover's algorithm. Unfortunately, this initial
state can only be prepared in the general case with the knowledge of the solution states.  However, 
we can overcome this by instead using a phase estimation procedure to solve the search problem instead, with a similar time scaling to the optimal case. This can lead to a reduction in the number of gates due to a simpler implementation of the Grover operator (see Supplementary Information). The phase estimation approach has the advantage that it can be applied in the general $ N, M $ case.   As amplitude amplification and phase estimation are typically considered to be different classes of quantum algorithm, it is interesting that in fact both approaches have a similar performance.  This suggests that phase estimation alone potentially gives a basis for performing both amplitude amplification and phase estimation based algorithms, which cover an extremely wide range of quantum algorithms known today.

The authors thank Jonathan Dowling for interesting discussions. This work is supported by the Shanghai Research Challenge Fund; New York University Global Seed Grants for Collaborative Research; National Natural Science Foundation of China (61571301); the Thousand Talents Program for Distinguished Young Scholars (D1210036A); and the NSFC Research Fund for International Young Scientists (11650110425); NYU-ECNU Institute of Physics at NYU Shanghai; the Science and Technology Commission of Shanghai Municipality (17ZR1443600); and the China Science and Technology Exchange Center (NGA-16-001).

.


\begin{thebibliography}{34}
\expandafter\ifx\csname natexlab\endcsname\relax\def\natexlab#1{#1}\fi
\expandafter\ifx\csname bibnamefont\endcsname\relax
  \def\bibnamefont#1{#1}\fi
\expandafter\ifx\csname bibfnamefont\endcsname\relax
  \def\bibfnamefont#1{#1}\fi
\expandafter\ifx\csname citenamefont\endcsname\relax
  \def\citenamefont#1{#1}\fi
\expandafter\ifx\csname url\endcsname\relax
  \def\url#1{\texttt{#1}}\fi
\expandafter\ifx\csname urlprefix\endcsname\relax\def\urlprefix{URL }\fi
\providecommand{\bibinfo}[2]{#2}
\providecommand{\eprint}[2][]{\url{#2}}

\bibitem[{\citenamefont{Grover}(1996)}]{grover1996}
\bibinfo{author}{\bibfnamefont{L.}~\bibnamefont{Grover}},
  \bibinfo{journal}{Proceedings of the 28th annual ACM} p. \bibinfo{pages}{212}
  (\bibinfo{year}{1996}).

\bibitem[{\citenamefont{Bennett et~al.}(1997)\citenamefont{Bennett, Bernstein,
  Brassard, and Vazirani}}]{Bennett1997}
\bibinfo{author}{\bibfnamefont{C.~H.} \bibnamefont{Bennett}},
  \bibinfo{author}{\bibfnamefont{E.}~\bibnamefont{Bernstein}},
  \bibinfo{author}{\bibfnamefont{G.}~\bibnamefont{Brassard}}, \bibnamefont{and}
  \bibinfo{author}{\bibfnamefont{U.}~\bibnamefont{Vazirani}},
  \bibinfo{journal}{SIAM J. Comput.} \textbf{\bibinfo{volume}{26}},
  \bibinfo{pages}{1510} (\bibinfo{year}{1997}).

\bibitem[{\citenamefont{F{\"u}rer}(2008)}]{Furer2008}
\bibinfo{author}{\bibfnamefont{M.}~\bibnamefont{F{\"u}rer}}, in
  \emph{\bibinfo{booktitle}{LATIN 2008: Theoretical Informatics: 8th Latin
  American Symposium, B{\'u}zios, Brazil, April 7-11, 2008. Proceedings}}
  (\bibinfo{publisher}{Berlin, Heidelberg}, \bibinfo{year}{2008}), pp.
  \bibinfo{pages}{784--792}.

\bibitem[{\citenamefont{Hsu}(2003)}]{Hsu03}
\bibinfo{author}{\bibfnamefont{L.-Y.} \bibnamefont{Hsu}},
  \bibinfo{journal}{Phys. Rev. A} \textbf{\bibinfo{volume}{68}},
  \bibinfo{pages}{022306} (\bibinfo{year}{2003}).

\bibitem[{\citenamefont{Hao et~al.}(2010)\citenamefont{Hao, Li, and
  Long}}]{Hao2010}
\bibinfo{author}{\bibfnamefont{L.}~\bibnamefont{Hao}},
  \bibinfo{author}{\bibfnamefont{J.}~\bibnamefont{Li}}, \bibnamefont{and}
  \bibinfo{author}{\bibfnamefont{G.}~\bibnamefont{Long}},
  \bibinfo{journal}{Science China Physics, Mechanics and Astronomy}
  \textbf{\bibinfo{volume}{53}}, \bibinfo{pages}{491} (\bibinfo{year}{2010}).

\bibitem[{\citenamefont{Magniez et~al.}(2007)\citenamefont{Magniez, Santha, and
  Szegedy}}]{Magniez07}
\bibinfo{author}{\bibfnamefont{F.}~\bibnamefont{Magniez}},
  \bibinfo{author}{\bibfnamefont{M.}~\bibnamefont{Santha}}, \bibnamefont{and}
  \bibinfo{author}{\bibfnamefont{M.}~\bibnamefont{Szegedy}},
  \bibinfo{journal}{SIAM Journal on Computing} \textbf{\bibinfo{volume}{37}},
  \bibinfo{pages}{413} (\bibinfo{year}{2007}).

\bibitem[{\citenamefont{Wang and Perkowski}(2011)}]{5954250}
\bibinfo{author}{\bibfnamefont{Y.}~\bibnamefont{Wang}} \bibnamefont{and}
  \bibinfo{author}{\bibfnamefont{M.}~\bibnamefont{Perkowski}}, in
  \emph{\bibinfo{booktitle}{2011 41st IEEE International Symposium on
  Multiple-Valued Logic}} (\bibinfo{year}{2011}), pp.
  \bibinfo{pages}{294--301}.

\bibitem[{\citenamefont{Zhang}(2005)}]{Chen05}
\bibinfo{author}{\bibfnamefont{C.-B.} \bibnamefont{Zhang}},
  \bibinfo{journal}{Journal of Optics B} \textbf{\bibinfo{volume}{7}}
  (\bibinfo{year}{2005}).

\bibitem[{\citenamefont{D{\"u}rr and H{\o}yer}(1996)}]{Durr96aquantum}
\bibinfo{author}{\bibfnamefont{C.}~\bibnamefont{D{\"u}rr}} \bibnamefont{and}
  \bibinfo{author}{\bibfnamefont{P.}~\bibnamefont{H{\o}yer}}
  (\bibinfo{year}{1996}), \eprint{quantph/9607014}.

\bibitem[{\citenamefont{Grover}(1997)}]{Grover1997}
\bibinfo{author}{\bibfnamefont{L.~K.} \bibnamefont{Grover}}
  (\bibinfo{year}{1997}), \eprint{quant-ph/9704012}.

\bibitem[{\citenamefont{{Ambainis}}(2005)}]{2005quant.ph.4012A}
\bibinfo{author}{\bibfnamefont{A.}~\bibnamefont{{Ambainis}}}
  (\bibinfo{year}{2005}), \eprint{quant-ph/0504012}.

\bibitem[{\citenamefont{Brassard et~al.}(1997)\citenamefont{Brassard, H{\o}yer,
  and Tapp}}]{Brassard1997}
\bibinfo{author}{\bibfnamefont{G.}~\bibnamefont{Brassard}},
  \bibinfo{author}{\bibfnamefont{P.}~\bibnamefont{H{\o}yer}}, \bibnamefont{and}
  \bibinfo{author}{\bibfnamefont{A.}~\bibnamefont{Tapp}}
  (\bibinfo{year}{1997}), \eprint{quant-ph/9705002}.

\bibitem[{\citenamefont{A{\"i}meur et~al.}(2013)\citenamefont{A{\"i}meur,
  Brassard, and Gambs}}]{Aimeur2013}
\bibinfo{author}{\bibfnamefont{E.}~\bibnamefont{A{\"i}meur}},
  \bibinfo{author}{\bibfnamefont{G.}~\bibnamefont{Brassard}}, \bibnamefont{and}
  \bibinfo{author}{\bibfnamefont{S.}~\bibnamefont{Gambs}},
  \bibinfo{journal}{Machine Learning} \textbf{\bibinfo{volume}{90}},
  \bibinfo{pages}{261} (\bibinfo{year}{2013}).

\bibitem[{\citenamefont{Nielsen and Chuang}(2000)}]{nielsen00}
\bibinfo{author}{\bibfnamefont{M.~A.} \bibnamefont{Nielsen}} \bibnamefont{and}
  \bibinfo{author}{\bibfnamefont{I.~L.} \bibnamefont{Chuang}},
  \emph{\bibinfo{title}{Quantum Computation and Quantum Information,}}
  (\bibinfo{publisher}{Cambridge University Press}, \bibinfo{address}{New
  York}, \bibinfo{year}{2000}).

\bibitem[{\citenamefont{Sadhukhan}(2012)}]{sadhukhan12}
\bibinfo{author}{\bibfnamefont{D.}~\bibnamefont{Sadhukhan}}, Ph.D. thesis,
  \bibinfo{school}{Indian Institute of Technology Bombay}
  (\bibinfo{year}{2012}).

\bibitem[{\citenamefont{Kato}(2005)}]{kato05}
\bibinfo{author}{\bibfnamefont{G.}~\bibnamefont{Kato}}, \bibinfo{journal}{Phys.
  Rev. A} \textbf{\bibinfo{volume}{72}}, \bibinfo{pages}{032319}
  (\bibinfo{year}{2005}).

\bibitem[{\citenamefont{Farhi and Gutmann}(1998)}]{PhysRevA.57.2403}
\bibinfo{author}{\bibfnamefont{E.}~\bibnamefont{Farhi}} \bibnamefont{and}
  \bibinfo{author}{\bibfnamefont{S.}~\bibnamefont{Gutmann}},
  \bibinfo{journal}{Phys. Rev. A} \textbf{\bibinfo{volume}{57}},
  \bibinfo{pages}{2403} (\bibinfo{year}{1998}).

\bibitem[{\citenamefont{Long}(2001)}]{PhysRevA.64.022307}
\bibinfo{author}{\bibfnamefont{G.~L.} \bibnamefont{Long}},
  \bibinfo{journal}{Phys. Rev. A} \textbf{\bibinfo{volume}{64}},
  \bibinfo{pages}{022307} (\bibinfo{year}{2001}).

\bibitem[{\citenamefont{Biron et~al.}(1998)\citenamefont{Biron, Biham, Biham,
  Grassl, and Lidar}}]{Biron:1998ic}
\bibinfo{author}{\bibfnamefont{D.}~\bibnamefont{Biron}},
  \bibinfo{author}{\bibfnamefont{O.}~\bibnamefont{Biham}},
  \bibinfo{author}{\bibfnamefont{E.}~\bibnamefont{Biham}},
  \bibinfo{author}{\bibfnamefont{M.}~\bibnamefont{Grassl}}, \bibnamefont{and}
  \bibinfo{author}{\bibfnamefont{D.~A.} \bibnamefont{Lidar}}, in
  \emph{\bibinfo{booktitle}{Lecture Notes in Computer Science}}
  (\bibinfo{year}{1998}), vol. \bibinfo{volume}{1509},
  \eprint{quant-ph/9801066}.

\bibitem[{\citenamefont{Grover}(2005)}]{grover2005fixed}
\bibinfo{author}{\bibfnamefont{L.~K.} \bibnamefont{Grover}},
  \bibinfo{journal}{Physical Review Letters} \textbf{\bibinfo{volume}{95}},
  \bibinfo{pages}{150501} (\bibinfo{year}{2005}).

\bibitem[{\citenamefont{Yoder et~al.}(2014)\citenamefont{Yoder, Low, and
  Chuang}}]{yoder2014fixed}
\bibinfo{author}{\bibfnamefont{T.~J.} \bibnamefont{Yoder}},
  \bibinfo{author}{\bibfnamefont{G.~H.} \bibnamefont{Low}}, \bibnamefont{and}
  \bibinfo{author}{\bibfnamefont{I.~L.} \bibnamefont{Chuang}},
  \bibinfo{journal}{Physical review letters} \textbf{\bibinfo{volume}{113}},
  \bibinfo{pages}{210501} (\bibinfo{year}{2014}).

\bibitem[{\citenamefont{Braunstein and van Loock}(2005)}]{braunstein05}
\bibinfo{author}{\bibfnamefont{S.}~\bibnamefont{Braunstein}} \bibnamefont{and}
  \bibinfo{author}{\bibfnamefont{P.}~\bibnamefont{van Loock}},
  \bibinfo{journal}{Rev. Mod. Phys.} \textbf{\bibinfo{volume}{77}},
  \bibinfo{pages}{513} (\bibinfo{year}{2005}).

\bibitem[{\citenamefont{Byrnes et~al.}(2015)\citenamefont{Byrnes, Rosseau,
  Khosla, Pyrkov, Thomasen, Mukai, Koyama, Abdelrahman, and
  Ilo-Okeke}}]{byrnes2014}
\bibinfo{author}{\bibfnamefont{T.}~\bibnamefont{Byrnes}},
  \bibinfo{author}{\bibfnamefont{D.}~\bibnamefont{Rosseau}},
  \bibinfo{author}{\bibfnamefont{M.}~\bibnamefont{Khosla}},
  \bibinfo{author}{\bibfnamefont{A.}~\bibnamefont{Pyrkov}},
  \bibinfo{author}{\bibfnamefont{A.}~\bibnamefont{Thomasen}},
  \bibinfo{author}{\bibfnamefont{T.}~\bibnamefont{Mukai}},
  \bibinfo{author}{\bibfnamefont{S.}~\bibnamefont{Koyama}},
  \bibinfo{author}{\bibfnamefont{A.}~\bibnamefont{Abdelrahman}},
  \bibnamefont{and}
  \bibinfo{author}{\bibfnamefont{E.}~\bibnamefont{Ilo-Okeke}},
  \bibinfo{journal}{Opt. Comm.} \textbf{\bibinfo{volume}{337}},
  \bibinfo{pages}{102} (\bibinfo{year}{2015}).

\bibitem[{\citenamefont{Pati et~al.}(2000)\citenamefont{Pati, Braunstein, and
  Lloyd}}]{pati00}
\bibinfo{author}{\bibfnamefont{A.~K.} \bibnamefont{Pati}},
  \bibinfo{author}{\bibfnamefont{S.~L.} \bibnamefont{Braunstein}},
  \bibnamefont{and} \bibinfo{author}{\bibfnamefont{S.}~\bibnamefont{Lloyd}}
  (\bibinfo{year}{2000}), \eprint{quantph/0002082}.

\bibitem[{\citenamefont{Byrnes et~al.}(2012)\citenamefont{Byrnes, Wen, and
  Yamamoto}}]{byrnes2012}
\bibinfo{author}{\bibfnamefont{T.}~\bibnamefont{Byrnes}},
  \bibinfo{author}{\bibfnamefont{K.}~\bibnamefont{Wen}}, \bibnamefont{and}
  \bibinfo{author}{\bibfnamefont{Y.}~\bibnamefont{Yamamoto}},
  \bibinfo{journal}{Phys. Rev. A} \textbf{\bibinfo{volume}{85}},
  \bibinfo{pages}{040306} (\bibinfo{year}{2012}).

\bibitem[{\citenamefont{Ermakov and Fung}(2002)}]{PhysRevA.66.042310}
\bibinfo{author}{\bibfnamefont{V.~L.} \bibnamefont{Ermakov}} \bibnamefont{and}
  \bibinfo{author}{\bibfnamefont{B.~M.} \bibnamefont{Fung}},
  \bibinfo{journal}{Phys. Rev. A} \textbf{\bibinfo{volume}{66}},
  \bibinfo{pages}{042310} (\bibinfo{year}{2002}).

\bibitem[{\citenamefont{Terhal and Smolin}(1998)}]{terhal98}
\bibinfo{author}{\bibfnamefont{B.}~\bibnamefont{Terhal}} \bibnamefont{and}
  \bibinfo{author}{\bibfnamefont{J.}~\bibnamefont{Smolin}},
  \bibinfo{journal}{Phys. Rev. A} \textbf{\bibinfo{volume}{58}},
  \bibinfo{pages}{1822} (\bibinfo{year}{1998}).

\bibitem[{\citenamefont{Poulin and Wocjan}(2009)}]{poulin09}
\bibinfo{author}{\bibfnamefont{D.}~\bibnamefont{Poulin}} \bibnamefont{and}
  \bibinfo{author}{\bibfnamefont{P.}~\bibnamefont{Wocjan}},
  \bibinfo{journal}{Phys. Rev. Lett.} \textbf{\bibinfo{volume}{103}},
  \bibinfo{pages}{220502} (\bibinfo{year}{2009}).

\bibitem[{\citenamefont{Farhi et~al.}(2014)\citenamefont{Farhi, Goldstone, and
  Gutmann}}]{farhi2014quantum}
\bibinfo{author}{\bibfnamefont{E.}~\bibnamefont{Farhi}},
  \bibinfo{author}{\bibfnamefont{J.}~\bibnamefont{Goldstone}},
  \bibnamefont{and} \bibinfo{author}{\bibfnamefont{S.}~\bibnamefont{Gutmann}},
  \bibinfo{journal}{arXiv preprint arXiv:1411.4028}  (\bibinfo{year}{2014}).

\bibitem[{\citenamefont{Brassard et~al.}(1998)\citenamefont{Brassard, H{\o}yer,
  and Tapp}}]{Brassard98}
\bibinfo{author}{\bibfnamefont{G.}~\bibnamefont{Brassard}},
  \bibinfo{author}{\bibfnamefont{P.}~\bibnamefont{H{\o}yer}}, \bibnamefont{and}
  \bibinfo{author}{\bibfnamefont{A.}~\bibnamefont{Tapp}}, in
  \emph{\bibinfo{booktitle}{Proceedings of 25th ICALP, Vol. 1443 of Lecture}}
  (\bibinfo{publisher}{Springer}, \bibinfo{year}{1998}), pp.
  \bibinfo{pages}{820--831}.

\bibitem[{\citenamefont{Mosca}(2001)}]{mosca2001counting}
\bibinfo{author}{\bibfnamefont{M.}~\bibnamefont{Mosca}},
  \bibinfo{journal}{Theoretical Computer Science}
  \textbf{\bibinfo{volume}{264}}, \bibinfo{pages}{139} (\bibinfo{year}{2001}).

\bibitem[{\citenamefont{Boyer et~al.}(1998)\citenamefont{Boyer, Brassard,
  H{\o}yer, and Tapp.}}]{boyer98}
\bibinfo{author}{\bibfnamefont{M.}~\bibnamefont{Boyer}},
  \bibinfo{author}{\bibfnamefont{G.}~\bibnamefont{Brassard}},
  \bibinfo{author}{\bibfnamefont{P.}~\bibnamefont{H{\o}yer}}, \bibnamefont{and}
  \bibinfo{author}{\bibfnamefont{A.}~\bibnamefont{Tapp.}},
  \bibinfo{journal}{Fortsch. Phys. – Prog. Phys}
  \textbf{\bibinfo{volume}{46}}, \bibinfo{pages}{493} (\bibinfo{year}{1998}).

\bibitem[{\citenamefont{Zalka}(1999)}]{zalka99}
\bibinfo{author}{\bibfnamefont{C.}~\bibnamefont{Zalka}},
  \bibinfo{journal}{Phys. Rev. A} \textbf{\bibinfo{volume}{60}},
  \bibinfo{pages}{2746} (\bibinfo{year}{1999}).

\end{thebibliography}

\end{document}